\documentclass[twocolumn, switch]{article} 

\usepackage{preprint}

\usepackage{amsmath, amsthm, amssymb, amsfonts}

\usepackage{natbib}
\usepackage[utf8]{inputenc}	
\usepackage[T1]{fontenc}	
\usepackage{xcolor}		
\usepackage[colorlinks = true,
            linkcolor = blue,
            urlcolor  = blue,
            citecolor = blue,
            anchorcolor = black]{hyperref}	
\usepackage{booktabs} 		
\usepackage{nicefrac}		
\usepackage{microtype}		
\usepackage{lineno}		
\usepackage{float}			

\usepackage{lipsum}		

\usepackage{newfloat}
\DeclareFloatingEnvironment[name={Supplementary Figure}]{suppfigure}
\usepackage{sidecap}
\sidecaptionvpos{figure}{c}
\usepackage[labelfont=bf, font=footnotesize, belowskip=-10pt]{caption}

\usepackage{titlesec}
\titlespacing\section{0pt}{12pt plus 3pt minus 3pt}{1pt plus 1pt minus 1pt}
\titlespacing\subsection{0pt}{10pt plus 3pt minus 3pt}{1pt plus 1pt minus 1pt}
\titlespacing\subsubsection{0pt}{8pt plus 3pt minus 3pt}{1pt plus 1pt minus 1pt}

\usepackage{tikz,xcolor,hyperref}

\title{MomentClosure.jl: automated moment closure approximations in Julia}

\usepackage{xwatermark}
\newwatermark[firstpage,color=gray!90,angle=0,scale=0.28, xpos=0in,ypos=-5in,width=15in]{*correspondence: \texttt{ramon.grima@ed.ac.uk} or \texttt{asukys@turing.ac.uk}}

\usepackage{authblk}

\newcommand*\samethanks[1][\value{footnote}]{\footnotemark[#1]}
\author[1,2\thanks{\tt{asmith@college.edu}}]{Augustinas Sukys}
\author[1\samethanks]{Ramon Grima}

\affil[1]{SynthSys, School of Biological Sciences, University of Edinburgh, Edinburgh EH9 3JD, UK}
\affil[2]{The Alan Turing Institute, London NW1 2DB, UK}

\begin{document}

\twocolumn[ 
  \begin{@twocolumnfalse} 
  
\maketitle

\begin{abstract}
MomentClosure.jl is a Julia package providing automated derivation of the time-evolution equations of the moments of molecule numbers for virtually any chemical reaction network using a wide range of moment closure approximations. It extends the capabilities of modelling stochastic biochemical systems in Julia and can be particularly useful when exact analytic solutions of the chemical master equation are unavailable and when Monte Carlo simulations are computationally expensive. 

MomentClosure.jl is freely accessible under the MIT license. Source code and documentation are available at \href{https://github.com/augustinas1/MomentClosure.jl}{https://github.com/augustinas1/MomentClosure.jl}
\end{abstract}
\vspace{0.35cm}

  \end{@twocolumnfalse} 
] 



\section{Introduction}

The stochastic dynamics of biochemical systems under well-mixed conditions are governed by the chemical master equation (CME). The CME cannot be solved analytically except for simple systems and its exact stochastic simulation \citep{Gillespie1977} can be computationally expensive, in turn motivating the development of more efficient approximation methods \citep{Schnoerr2017}. 

One approach is to approximate the whole probability distribution solution of the CME in terms of its first few moments: starting from the CME, we can derive a set of ordinary differential equations (ODEs) describing the time-evolution of moments for the molecule numbers of each species in a system, e.g., means and (co)variances. However, if a chemical reaction network is non-linear, we end up with an infinite hierarchy of coupled moment equations that cannot be solved directly as each moment will depend on higher order moments. Nevertheless, this problem can be tackled using one of many moment closure approximations (MAs) that express all moments above a certain order in terms of lower order moments using various (usually distributional) assumptions, effectively closing the hierarchy and enabling a numerical solution \citep{Lakatos2015, Schnoerr2015, Soltani2015}. 

Deriving the moment equations and applying MAs manually can be a cumbersome and error prone process, especially when large systems and high-order MAs are considered. For this reason, a number of software tools have been developed over the years, allowing automatic derivation of the closed moment equations for a specified chemical reaction network \citep{StochDynTools, MomentClosureGillespie, MFKpackage, Schnoerr2015, MEANS, CERENA}. However, these packages are either outdated and unmaintained, offer limited functionality (implementing only few types of MAs and restricted to mass action reactions) or require proprietary expensive software, limiting the potential user base.

We present MomentClosure.jl, the first Julia package to automatically derive the closed time-evolution equations of moments up to an arbitrary order for any chemical reaction network supporting both non-polynomial and time-dependent propensity functions, and implementing a variety of MAs commonly used in stochastic biochemical kinetics. A comprehensive review of MAs and tutorials on using the software can be found at \href{https://github.com/augustinas1/MomentClosure.jl}{https://github.com/augustinas1/MomentClosure.jl}.

\section{Materials and methods}

\begin{figure}
\centerline{\includegraphics[width=1.0\linewidth]{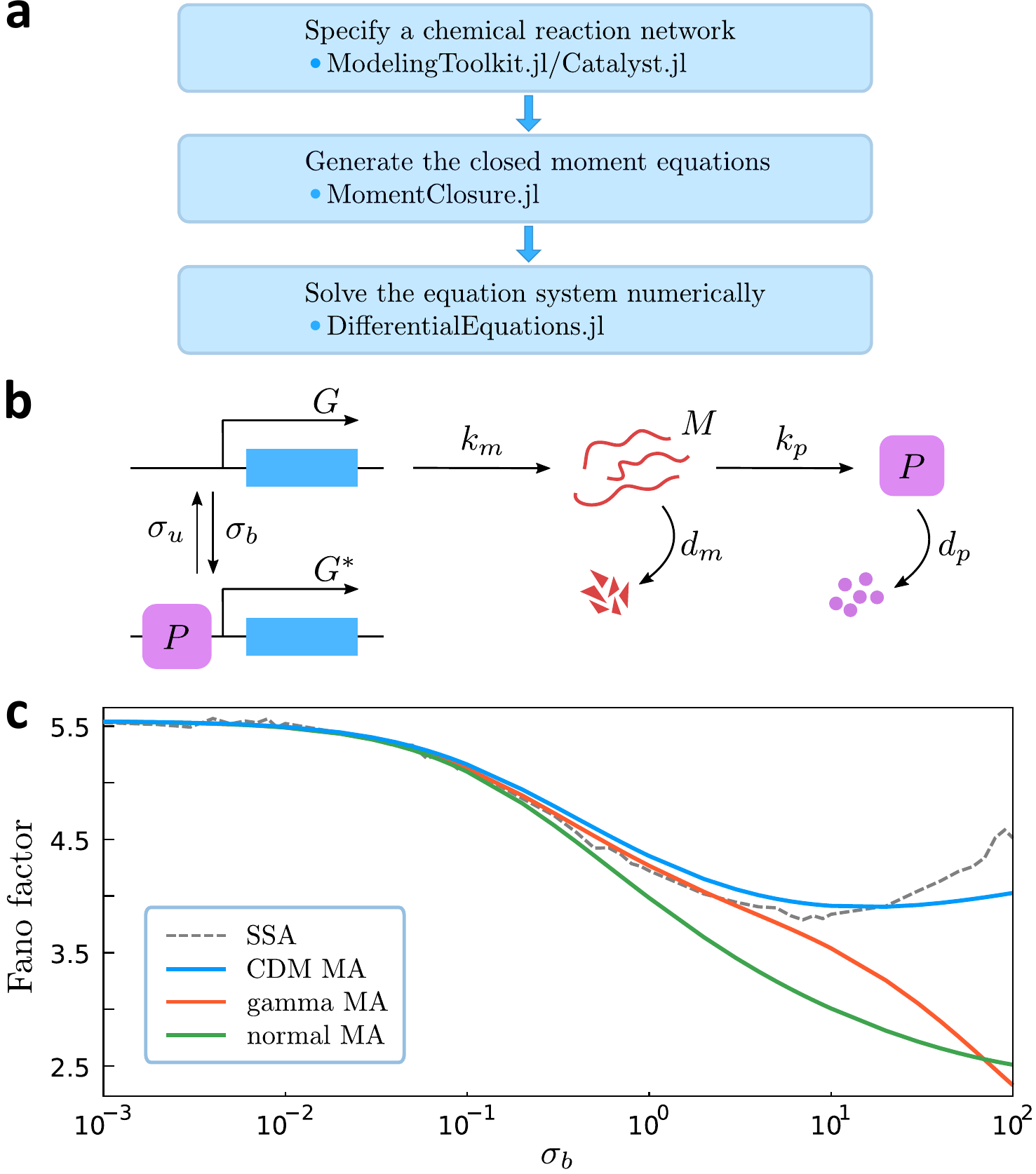}}
\caption{(\textbf{a}) The general workflow of moment-based modelling in Julia using MomentClosure.jl and related packages. (\textbf{b}) Model of a negative auto-regulative genetic feedback loop. If a gene is active ($G$), an mRNA molecule ($M$) is produced with rate $k_m$ which can be subsequently translated into proteins ($P$) with rate $k_p$ (both degrade with rates $d_m$ and $d_p$ respectively). The negative feedback is introduced via protein binding to the gene with rate $\sigma_b$, switching the promoter OFF ($G^*$) and preventing the transcription (in contrast, switching ON occurs with rate $\sigma_u$). (\textbf{c}) Fano factor of the steady-state protein number as a function of $\sigma_b$ where we have truncated the moment hierarchy at the second order using normal, gamma and conditional derivative matching (CDM) MAs, and compared the results to the true values predicted by the stochastic simulation algorithm \citep{Gillespie1977} (averaged over $10^5$ realisations). The initial condition is zero protein and mRNA molecules (set to $0.001$ for MAs to ensure numerical stability) in state $G$, and the parameters are fixed as $\sigma_u = 10, k_m = 3, k_p = 50, d_m = 10$ and $d_p = 1$.} \label{fig:fig1} \end{figure}

In what follows, we discuss the main implementation details of MomentClosure.jl and its integration within the broader Julia package ecosystem enabling a streamlined moment-based modelling workflow in which we can easily define a biochemical system, generate the corresponding moment equations using MAs and solve the resulting system of ODEs numerically. In Figure~\ref{fig:fig1}, we summarise the workflow and compare the accuracy of different MAs applied to a simple stochastic model of an auto-regulatory genetic feedback loop.

\subsection{Model definition}

Modelling of chemical reaction networks in Julia is made easy by Catalyst.jl (\href{https://github.com/SciML/Catalyst.jl}{https://github.com/SciML/Catalyst.jl}) that leverages a powerful symbolic-numeric modelling framework provided by ModelingToolkit.jl \citep{ModelingToolkit}: a model can be constructed by simply specifying the reaction stoichiometry and the propensities using symbolic variables. MomentClosure.jl is fully compatible with models defined through the two packages, allowing systems containing any number of molecular species and reactions with any type of smooth propensity functions. The only assumption made is that reactions occur in a single compartment of fixed volume. We note that networks involving multiple dynamically interacting compartments can be considered using Compartor \citep{Compartor}, albeit the software is restricted to reactions with polynomial rate laws and supports only mean-field and gamma MAs.

\subsection{Moment equations}

Using MomentClosure.jl, we can automatically obtain a system of ODEs describing the time evolution of moments up to any order. Internally, the \emph{raw} moment equations are derived in a straightforward manner from the CME when the kinetics of a system are governed by the law of mass action. The derivation becomes more involved if the propensity functions take a non-polynomial form: here we adopt a general moment expansion framework based on Taylor-expanding the propensities around the mean, allowing us to obtain equations for the means and higher order \emph{central} moments, as first formulated by \citep{Lee2013} and \citep{Ale2013}.

The moment equations generated up to the specified order $m$ can then be decoupled from all the higher order moments they depend on using one of the implemented MAs, i.e, expressing the moments above order $m$ in terms of $m^{\text{th}}$ and lower order moments using MA-specific assumptions. MomentClosure.jl currently supports zero, normal, Poisson, log-normal, gamma, derivative matching, conditional Gaussian and conditional derivative matching closures \citep{Lakatos2015, Soltani2015, Schnoerr2015}. We note that the conditional MAs are not available in other packages and have been found to be particularly effective in modelling gene networks with promoter switching dynamics \citep{Soltani2015, Cao2019}.

Finally, the closed moment equations can be solved numerically using any high-performance ODE solver implemented in DifferentialEquations.jl \citep{Rackauckas2017}, which also provides a number of numerical analysis and parameter estimation tools enabling even further study of the resulting ODE system.

\section{Conclusion}

MomentClosure.jl provides automated moment equation generation and closure approximations up to any desired expansion order. It is easily applicable to chemical reaction networks of any size containing reactions with smooth linear and nonlinear propensity functions. Moreover, utilising the popular ModellingToolkit.jl and DifferentialEquations.jl packages, MomentClosure.jl makes the stochastic modelling of biochemical reaction kinetics using MAs efficient and accessible for Julia newcomers and experts alike.

\section*{Funding}

This work was supported by the the Alan Turing Institute Doctoral Studentship for A.S (EPSRC grant EP/N510129/1), and a Leverhulme Trust grant
(Grant No. RPG-2018-423) for R.G. \\

\noindent\emph{Conflict of Interest}: none declared.



\bibliography{references}


\end{document}